\documentclass{jaa}

\usepackage{graphicx}
\usepackage{color}

\begin{document}

\title{Standing Shocks in Magnetized Dissipative Accretion
Flow around Black Holes}


\author{Biplob Sarkar\textsuperscript{*}, Santabrata Das}
\affilOne{Indian Institute of Technology Guwahati, Guwahati 781039, Assam, India}


\twocolumn[{

\maketitle

\corres{biplob@iitg.ernet.in}

\msinfo{xx xxx 201x}{xx xxx 201x}{xx xxx 201x}

\begin{abstract}
We explore the global structure of the accretion flow around a Schwarzschild black hole 
where the accretion disc is threaded by toroidal magnetic fields. The accretion flow is optically 
thin and advection dominated. Synchrotron radiation is considered to be the active cooling 
mechanism in the flow. With this, we obtain the global transonic accretion solutions and show 
that centrifugal barrier in the rotating magnetized accretion flow causes a discontinuous 
transition of the flow variables in the form of shock waves. The shock properties and the 
dynamics of the post-shock corona are affected by the flow parameters such as viscosity, 
cooling rate and strength of the magnetic fields. The shock properties are investigated against 
these flow parameters. We further show that for given set of boundary parameters at the 
outer edge of the disc, accretion flow around a black hole admits shock when the flow parameters 
are tuned for a considerable range.  
\end{abstract}


}]


\doinum{12.3456/s78910-011-012-3}
\artcitid{\#\#\#\#}
\volnum{123}
\year{2017}
\pgrange{23--25}
\setcounter{page}{23}
\lp{25}

\section{Introduction}
Magnetic fields are abundant in the astrophysical environment and due to their presence in accretion 
discs around black holes, these systems exhibit many interesting features. Ichimaru (1977) pointed out 
the significance of magnetic fields on the transition between an optically thin hot disc and an optically 
thick cool disc. He argued that the magnetic pressure ($P_{\rm mag}$) cannot exceed the gas pressure ($P_{\rm gas}$) 
since the magnetic flux will escape from the disc due to buoyancy. Magnetohydrodynamic 
(MHD) simulations of buoyant escape of magnetic flux as a result of Parker instability (Parker 1966) was 
performed by Shibata {\em et al.} (1990). This simulation showed that when magnetic pressure is 
dominant in the disc, the disc can survive in the low-$\beta$ state ($\beta$ = $P_{\rm gas}$/$P_{\rm mag} < 1$) 
since the growth rate of Parker instability suffers a decline in the low-$\beta$ state because of magnetic tension. 
These findings point to the fact that magnetized accretion disc systems are astrophysically viable.

The consequences of large scale ordered magnetic fields in accretion disc theories are frequently 
investigated in two categories. In one category, the global magnetic field is considered where both the poloidal and 
toroidal components of the ordered magnetic field are present. In the other category, the accretion disc is threaded 
only by the toroidal field. The latter is justified since an accretion disc is rotationally dominated and we expect 
the magnetic fields in the disc to be primarily toroidal in nature. Toroidal field is generated in the disc by the effect 
of differential rotation on the originally poloidal field lines linking layers rotating at different rates (Papaloizou \& Terquem 
1997). Thus in this article, we consider the accretion disc to be threaded by toroidal magnetic fields.

Numerical MHD simulations by Hirose {\em et al.} (2004) showed that the toroidal component of the magnetic 
field governs the main body of the accretion flow, principally in its inner region, whereas regions close to the poles 
are predominantly governed by a poloidal magnetic field, which is chiefly in the vertical direction. The poloidal component 
of magnetic field is of primary importance in accretion discs to account for the origin of outflows through the magnetocentrifugal 
force (Ustyugova {\em et al.} 1999, Spruit \& Uzdensky 2005). However, global three-dimensional MHD simulations (e.g., Hawley 2001; Kato 
{\em et al.} 2004) have showed that inside the disc, the poloidal component of magnetic field is weak compared to the azimuthal component. 
Since the toroidal component of the field is significantly larger than the radial and vertical components, when the toroidal field 
buoyantly escapes from the disc, the so-called `hoop stress' would greatly assist in collimating the flow (Pudritz \& Norman 
1986, Lebedev {\em et al.} 2005, Singh \& Chakrabarti 2011). However, investigation 
of the various initial magnetic field geometries by many authors (e.g., Hawley \& Krolik 2002, Igumenshchev {\em et al.} 2003, Beckwith 
{\em et al.} 2008 and McKinney \& Blandford 2009) have shown that the evolution of models with exclusively toroidal initial field is very 
moderate as compared to those with a poloidal initial field. This is so since the former have neither a vertical field to begin with, 
that is required for the linear MRI, nor a radial field, that is necessary for field amplification through shear. Inflow starts only 
after the MRI has generated turbulence of adequate amplitude (Hawley \& Krolik 2002), which occurs considerably 
later when the field is toroidal to begin with.

In the context of observational aspects, magnetic fields have been proposed to play an important role for the generation/collimation of jets
in black hole systems. It has been shown by exhaustive magnetohydrodynamic simulations of accretion disc around spinning black holes that
windy hot materials (i.e. corona) escape from the inner part of the disc as jets (Koide {\em et al.} 2002; McKinney \& Gammie 2004;
De Villiers {\em et al.} 2005). In the vicinity of the horizon, spin of the black hole drags the space-time geometry and the magnetic field lines
are twisted consequently transporting energy from the black hole along the field lines. Blandford \& Znajek (1977) showed that this 
mechanism has the potential to produce truly relativistic jets. Furthermore, results reported by Oda {\em et al.} (2010) indicated the significance of the
magnetically supported disc where it was shown that the model can provide a satisfactory explanation for the bright/hard state observed during the
bright hard-to-soft transition of galactic black hole candidates. Identification of such state transition in X-ray observations have been reported by 
Gierli\'nski \& Newton (2006).

Magnetic fields in accretion discs play the dual role. In one hand, they contribute to 
angular-momentum transport thus making accretion possible and on the other hand, dissipation 
of magnetic energy contributes to the heating of the disc (Hirose {\em et al.} 2006; Krolik {\em et al.} 2007). 
The conventional model of accretion discs (e.g., Shakura \& Sunyaev 1973 (SS73)) appeal to the 
phenomenological $\alpha-$viscosity. Yet, the exact physical mechanism that allows adequate angular 
momentum transport to explain the activities of accretion-powered
sources, such as dwarf nova, remains inconclusive. Balbus and Hawley (1991) revealed 
the relevance of the magneto-rotational instability (MRI) in accretion discs. MRI 
can successfuly excite and maintain magnetic turbulence. The Maxwell stress created by the turbulent 
magnetic fields effectively transports angular momentum enabling disc material to 
accrete. In the quasi-steady state, the average ratio of the Maxwell stress to the gas
pressure ($\alpha_{\rm SS}$) which correlates with the $\alpha$-parameter (value 
of the stress-to-pressure ratio) in the conventional accretion disc models (SS73), is $0.01-0.1$ 
(e.g., Hawley 2000; Hawley \& Krolik 2001; Machida \& Matsumoto 2003).

The above findings have also been validated by numerical simulations. Machida {\em et al.} (2006) performed 
global three-dimensional MHD simulations of black hole (BH) accretion discs. The 
simulation was started considering a radiatively inefficient torus threaded by weak toroidal magnetic 
fields. With the growth of MRI, an optically thin, hot accretion disc is formed by the efficient 
transport of angular momentum. When the density of the accretion disc exceeds the critical density, 
a cooling instability takes place in the disc and gas pressure decreases due to cooling. Hence, disc 
shrinks in vertical direction while attempting to almost conserve toroidal magnetic flux. This leads magnetic 
fields to amplify due to flux conservation and eventually $P_{\rm mag}$ exceeds $P_{\rm gas}$. With this, 
a magnetically supported disc is formed. When $P_{\rm mag}$ becomes dominant, disc stops shrinking 
in vertical direction because the magnetic pressure supports the disc. Finally, an optically thin, 
radiatively inefficient, hot, high-$\beta$ disc undergoes transition to an optically thin, radiatively 
efficient, cool, low-$\beta$ disc, except in the plunging region (e.g., Machida {\em et al.} 2006; Oda {\em et al.} 2010).

In an accretion flow around a black hole, radial velocity of the flow reaches at speed of light at the 
event horizon while sound speed remains lesser. At large distance, the flow may be at rest while 
still having some temperature (nonzero sound speed). Therefore, flow is supersonic at the event horizon 
and subsonic at a large distance. Hence flow must pass through at least one sonic point and presumably 
more, in presence of even a small angular momentum. In presence of multiple sonic points, flow is richer in 
topological properties and may contain dynamically important shock waves. Since matter must enter the black 
hole supersonically, the flow must therefore be sub-Keplerian close to the black hole horizon even in the 
presence of heating and cooling processes (Chakrabarti 1996; Chakrabarti 1990). 

The advective accretion process around a black hole is the competition between gravitational ($F_{\rm GR}$) 
and centrifugal ($F_{\rm CEN}$) forces. The slowed down inner part of the disc acts as effective boundary layer 
around the black holes (Chakrabarti 1996; Chakrabarti 1999; Das {\em et al.} 2001b; Chakrabarti \& Das 2004; Das 2007; 
Chattopadhyay \& Chakrabarti 2011). Therefore, centrifugal barrier triggers the formation of 
shock. Following the Second Law of Thermodynamics, shock is preferred in accretion solution for the sake of 
choosing the high-entropy solution when possible (Fukue 1987; Chakrabarti 1989; Lu \& Yuan 1998; Fukumura \& Tsuruta 2004)

Based on the above considerations, we investigate a single-temperature, magneto-fluid model for accretion 
disc around a stationary black hole. We consider synchrotron cooling mechanism as energy dissipation
process following Shapiro and Teukolsky (1983). The sonic point analysis and Rankine-Hugoniot shock conditions (Landau 
\& Lifshitz 1959) are employed following the treatment of Sarkar \& Das (2015, 2016); Sarkar em {\em et al.} (2018). 
With these assumptions, we show the presence of accretion solutions passing through the outer sonic point and 
show that these type of solutions may possess shock waves depending on the fulfillment of the shock conditions. 

In Sarkar \& Das (2016) the authors investigated magnetized accretion discs around non-rotating black holes 
employing the radial variation of magnetic flux in presence of bremsstrahlung cooling process. The authors 
showed that such discs harbour shocks for a wide range of flow parameters. However, the strength of the 
magnetic field was moderate in this work. Very recently, Sarkar {\em et al.} (2018) explored the possibility of 
shocks in magnetically supported accretion discs around stationary black holes in the presence of 
synchrotron cooling. Since accretion discs around stellar mass black holes are threaded by significant 
amount of magnetic fields, the authors focussed their interest on these systems. The present paper is 
a follow-up work of Sarkar {\em et al.} (2018). Nonetheless, the present paper further strengthens the 
claims made by Sarkar {\em et al.} (2018). In particular, the present work considers about the global 
transonic solution with shock, especially passing through an outer sonic point by considering the radial 
structure of the magnetic field. Such solutions passing through outer sonic point are highly important to 
explain many aspects of BH accretion properties, but they have been mostly ignored in the literature. We 
have provided a detailed analysis of the radial variation of the structure and strength of large-scale 
magnetic field in the disc. Many works in the literature have considered only the random or stochastic 
magnetic field in the disc to study transonic accretion solutions with or without shocks and ignored the 
large-scale field (Narayan \& Yi 1995; Yuan 2001, Mandal \& Chakrabarti 2005, Das 2007, Rajesh \& Mukhopadhyay 2010). 
The present paper aims to emphasize that shocks in accretion flows are present even when large-scale 
toroidal fields thread the disc. For a chosen set of outer edge parameters, we have shown that the magnetic 
field strength can reach $> 10^{6.5} G$ in the inner region of the accretion disc. The electron temperature 
in the disc regulates the cutoff in X-ray spectra originating from the disc. Oda {\em et al.} 2012 reported that 
the electron temperature ($T_e$) in the low$-\beta$ solutions is lower ($T_e \sim 10^8-10^{9.5} K$) than that in 
advection-dominated accretion flow (ADAF) ($T_e \geq 10^{9.5} K$). Our results are also in consonance with this finding. Also the estimated 
vertical optical depth in the disc is shown to remain $\ll 1$ even when magnetic pressure is dominant in the 
disc and this essentially indicates that the possibility of escaping hard radiations from the disc is quite high. The paper 
is organized as follows. In Section 2, we present the model assumptions and governing equations. In Section 3, we obtain the global 
accretion solutions with and without shock, shock properties, and the critical limits of accretion rate. Finally in Section 4, we 
present the summary and concluding remarks. 
  
\section{Basic model and Governing Equations}

Results of numerical simulations of global and local MHD accretion flow around black holes have shown that magnetic fields inside the disc are turbulent and the azimuthal component predominates (e.g., Machida {\em et al.} 2006; Johansen \& Levin 2008). Resting on the findings of these simulations, we decomposed the magnetic fields into mean fields and fluctuating fields. 
In this process, we considered the mean fields as ${\bf{B}} = (0,<B_{\phi}>,0)$, while the fluctuating 
fields are expressed as $\delta {\bf{B}} = (\delta B_r,\delta B_{\phi},\delta B_z)$. Here, $<>$ indicates the 
azimuthal average. We consider that the fluctuating components vanish upon azimuthally averaging, $<\delta {\bf{B}}> = 0$. Also, we assume that the radial and vertical components of the magnetic field are insignificant as compared to the azimuthal component, $|<B_{\phi}> + \delta B_{\phi}|\gg|\delta B_r|$ and $|\delta B_z|$. Essentailly, this provides the azimuthally averaged magnetic field as $<{\bf{B}}>=<B_{\phi}>\hat{\phi}$ (Oda {\em et al.} 2007).

\subsection{Governing Equations}
In the present model, we consider a steady, axis-symmetric, geometrically thin, viscous disc around a
Schwarzschild black hole of mass, $M_{\rm BH}$. We use the Geometric unit system as $2G = M_{\rm BH} = c =1$, where, $G$ 
is the universal Gravitational constant and $c$ is the speed of light. In this unit system, length, time 
and velocity are measured in unit of ${2GM_{\rm BH}}/{c^2}$, ${2GM_{\rm BH}}/{c^3}$ and $c$, respectively. We 
assume matter being accreted through the equatorial plane of the black hole. Accordingly, we adopt cylindrical 
polar coordinates ($x,\phi,z$) with the black hole located at the origin.

The set of governing equations describing the accretion flow around black hole in the steady state are given by:\\
(a) Radial momentum equation:
$$
{\upsilon\frac{d\upsilon}{dx} + \frac{1}{\rho}\frac{dP}{dx} - \frac{\lambda^2(x)}{x^3} + \frac{d\Psi}{dx} + \frac{\left<B_{\phi} ^2\right>}{4\pi x \rho} = 0}.
\eqno(1)
$$ where, $\upsilon$ is the radial velocity, $\rho$ is the density and $\lambda$ is the specific angular momentum of 
the flow, respectively. $P$ denotes the total pressure in the disc. We consider $P = p_{\rm gas} + p_{\rm mag}$, 
where $p_{\rm gas}$ is the gas pressure and $p_{\rm mag}$ is the magnetic pressure of the flow, respectively. 
The gas pressure inside the disc is given by, $p_{\rm gas} = R \rho T/\mu$, where, $R$ is the gas constant, $\mu$ 
is the mean molecular weight assumed to be $0.5$ for fully ionized hydrogen and $T$ is the temperature. The 
azimuthally averaged magnetic pressure is given by, $p_{\rm mag} = <B_{\phi}^2>/8\pi$. We define plasma 
$\beta = p_{\rm gas}/p_{\rm mag}$ and this provides $P = p_{\rm gas} (1 + 1/\beta)$. The effect of space-time geometry 
around a stationary black hole has been approximated by using the pseudo-Newtonian potential (Paczynski \& Wiita (1980)), 
given by,

$$
\Psi = -\frac{1}{2(x -1)},
\eqno(2)
$$\\
(b) Mass Conservation:
$$
\dot{M}=2\pi x\Sigma \upsilon,
\eqno(3)
$$ where, $\dot{M}$ is mass accretion rate which is constant globally. $\Sigma$ 
represents the vertically integrated density of flow (Matsumoto {\em et al.} 1984).\\
(c) Azimuthal momentum equation:\\
$$
\upsilon\frac{d\lambda(x)}{dx}+\frac{1}{\Sigma x}\frac{d}{dx}(x^2T_{x\phi}) = 0,
\eqno(4)
$$
where, we consider that the vertically integrated total stress to be dominated by the $x\phi$ component of 
the Maxwell stress $T_{x\phi}$. Following Machida {\em et al.} (2006) and for an advective flow with significant radial velocity, 
we estimate $T_{x\phi}$ as (Chakrabarti \& Das 2004)
$$
T_{x\phi} = {\frac{<B_{x}B_{\phi}>}{4\pi}}h = -\alpha _{B}(W + \Sigma \upsilon^2),
\eqno(5)  
$$ where, $h$ denotes the the half thickness of the disc, $\alpha_B$ (ratio of Maxwell stress to the 
total pressure) is the constant of proportionality and $W$ is the vertically integrated pressure (Matsumoto 
{\em et al.} 1984). In the present work, $\alpha_B$ is treated as a parameter based on the seminal 
work of SS73. When the radial velocity is unimportant, such as for a Keplerian flow, Eq. (5) 
subsequently reduces to the original prescription of `$\alpha-$model' (SS73).

Assuming the flow to be in vertical hydrostatic equilibrium, $h$ is given by,
$$
h = \sqrt{\frac{2}{\gamma}} a x^{1/2} (x - 1),
\eqno(6)
$$
where, the adiabatic sound speed is defined as $a=\sqrt {\gamma P/\rho}$, where $\gamma$ is the adiabatic index. 
We assume $\gamma$ to remain constant globally and adopt the canonical value of $\gamma=4/3$ 
in the subsequent analysis.\\
(d) The entropy generation equation:
$$
\Sigma \upsilon T \frac {ds}{dx}=\frac{h\upsilon}{\gamma-1}
\left(\frac{dp_{\rm gas}}{dx} -\frac{\gamma p_{\rm gas}}{\rho}\frac{d\rho}{dx}\right)=Q^- - Q^+,
\eqno(7)
$$
where, $s$ and $T$ represent the specific entropy and the local temperature of the flow, respectively. 
In the right hand side, $Q^+$ and $Q^-$ denote the vertically integrated heating and cooling rates. 
Numerical simulations show that heating of the flow arises due to the thermalization of magnetic 
energy through the magnetic reconnection mechanism (Hirose {\em et al.} 2006; Machida {\em et al.} 2006) and therefore, 
expressed as
$$
Q^{+} = \frac{<B_{x}B_{\phi}>}{4\pi} xh \frac{d\Omega}{dx} = -\alpha _{B}(W + \Sigma \upsilon^2) x \frac{d\Omega}{dx},
\eqno(8) 
$$ where, $\Omega$ denotes the local angular velocity of the flow.

In the present analysis, since magnetic fields play an important role in the accretion disc, it is 
evident that electrons should primarily emit synchrotron radiation. Thus considering synchrotron process 
as the effective radiative cooling mechanism, the cooling rate of the flow is given 
by (Shapiro \& Teukolsky 1983),
$$
Q^-= \frac{Sa^5 \rho h}{\upsilon x^{3/2}(x-1)} \frac{\beta^2}{(1+\beta)^3},
\eqno(9)
$$ with,
$$
S= 1.048 \times 10^{18} \frac{ {\dot m} \mu^2 e^4}{I_n m_e^3\gamma^{5/2}}
\frac{1}{2GM_{\odot}c^3},
$$
where, $e$ and $m_e$ denote the charge and mass of the electron, respectively. Also, $k_B$ is the Boltzmann constant, 
$I_n = (2^n n!)^2/(2n + 1)!$ and $n = 1/(\gamma - 1)$ is the polytropic constant. The electron temperature is estimated using the relation 
$T_e = (\sqrt{m_e/m_p})T_p$ (Chattopadhyay \& Chakrabarti 2002), where any coupling between the ions and electrons 
have been ignored. Furthermore, in subsequent sections, $\dot m$ represents the accretion rate measured in units of 
Eddington rate ($\dot{M}_{\rm Edd} = 1.39 \times 10^{17} \times M_{\rm BH}/M_{\odot}$ $gs^{-1}$).\\
(e) Radial advection of the toroidal magnetic flux:

To account for the radial advection of the toroidal magnetic flux, we consider the induction 
equation and it is expressed as,

$$
\frac {\partial <B_{\phi}>\hat{\phi}}{\partial t} = {\bf \nabla} \times
\left({\vec{\upsilon}} \times <B_{\phi}>\hat{\phi} -{\frac{4\pi}{c}}\eta {\vec{j}}\right),
\eqno(10)
$$
where, $\vec{\upsilon}$ is the velocity vector, $\eta$ is the resistivity, and 
${\vec{j}} = c({\nabla} \times <B_{\phi}>\hat{\phi})/4\pi$ 
denotes the current density. Here, Eq. (10) is azimuthally averaged. On account of very 
large length scale of accretion disc, the Reynolds number ($R_m$) is very large rendering the 
magnetic-diffusion term insignificant, thus, it is neglected. Further we also drop the 
dynamo term for the present study. In the steady state, the resultant equation is then 
vertically averaged considering that the azimuthally averaged toroidal magnetic fields vanish at 
the disc surface. As an outcome, we get the advection rate of the toroidal magnetic 
flux as (Oda {\em et al.} 2007),

$$
\dot{\Phi} = - \sqrt{4\pi}\upsilon h {B}_{0} (x) ,
\eqno(11)
$$
where,
\begin{eqnarray*}
{B}_{0} (x) && = \langle {B}_{\phi} \rangle \left(x; z = 0\right)  \nonumber \\
&& = 2^{5/4}{\pi}^{1/4}(R T/\mu)^{1/2}{\Sigma}^{1/2}h^{-1/2}{\beta}^{-1/2}
\end{eqnarray*}
denotes the azimuthally averaged toroidal magnetic field restrained in the disc equatorial plane. 
If the contribution from the dynamo term and the magnetic diffusion term do not cancel each 
other in the whole region, $\dot{\Phi}$ is expected to vary in the 
radial direction. Global three-dimensional MHD simulation by Machida {\em et al.} (2006) indicates 
that in the quasi-steady state, as a result of the aforementioned processes, the magnetic flux 
advection rate varies as, $\dot{\Phi} \propto 1/x$. Explicit computation of the magnetic diffusion 
term and the dynamo term are hard from the local quantities. Hence, by introducing a parameter, 
$\zeta$, the dependence of $\dot{\Phi}$ on $x$ is parameterized (Oda {\em et al.} 2007) and is given by,

$$
\dot{\Phi}\left(x; \zeta, \dot{M}\right) \equiv \dot{\Phi}_{\rm edge}(\dot{M})\left(\frac{x}{x_{\rm edge}} \right)^{-\zeta},
\eqno(12)
$$
where, $\dot{\Phi}_{\rm edge}$ denotes the advection rate of the toroidal magnetic flux at the outer edge 
of the disc ($x_{\rm edge}$). When $\zeta = 0$, it implies the conservation of  magnetic flux in the radial 
direction, and when $\zeta > 0$, the magnetic flux increases as the accreting matter advances towards the black 
hole horizon. In this work, for the sake of representation, we fix $\zeta = 1$, and assume $\zeta$ to 
remain constant all throughout unless stated otherwise.

\subsection{Sonic Point Analysis}

We aim to obtain a global accretion solution in the steady-state, where inflowing matter 
from a large distance can smoothly accrete inwards before going in to the black hole. 
Additionally, in order to preserve the inner boundary condition imposed by the event horizon, 
the flow must necessarily become transonic. On the basis of the above observations, the general 
nature of the sonic points is inferred from simultaneous solution of equations (1), (3), (4), 
(7), (11) and (12) which is expressed as,

$$
\frac {d\upsilon}{dx}=\frac{N}{D},
\eqno(13)
$$
where, the expressions of numerator $N$ and the denominator $D$ are same as in Sarkar {\em et al.} (2018). 
Moreover, the gradient of sound speed, angular momentum and plasma $\beta$, can also be referred to from Sarkar {\em et al.} (2018).

It is previously pointed out that since the flow trajectory must be smooth everywhere along the streamline, it 
demands that the velocity gradient must be real and finite always. {However, there exists some points between 
the outer edge of the disc and the horizon, where the denominator ($D$) may vanish. In order 
for the solution to be continuous, the point where $D$ goes to zero, $N$ must also simultaneously 
tends to zero there. Such a location, where $N$ and $D$ vanish simultaneously is known as a sonic 
point ($x_c$). So, we attain two conditions at the sonic point as $N = 0$ and $D = 0$. Equating $D$ to 
zero, we attain the expression of Mach number ($M = \upsilon/a$) at the sonic point as,

$$
M(x_c) =\sqrt {\frac{-m_2 - \sqrt{m^2_2-4m_1 m_3}}{2m_1}},
\eqno(14)
$$
where,
$$ 
m_1=2\alpha^2_{B} I_n \gamma^2(1 + \beta_c)(\gamma-1)(2g-1) - \gamma^2(3 + (\gamma+1)\beta_c)
$$
$$
m_2=2\gamma(2 + \gamma\beta_c) + 4\alpha^2_{B} I_n \gamma g (1 + \beta_c)(g-1)(\gamma-1)
$$ 
$$
m_3=-2\alpha^2_{B} I_n g^2 (1 + \beta_c)(\gamma-1)
$$ 

Employing the other sonic point condition $N = 0$, we attain an algebraic equation of the sound speed at 
$x_c$ which is given by,

$$
{\mathcal A}a^3(x_c) + {\mathcal B}a^2(x_c) + {\mathcal C}a(x_c) +{\mathcal D}= 0 ,
\eqno(15)
$$
where,
$$
{\mathcal A}=\frac{S}{x_c^{3/2}(x_c-1)}\frac{\beta_c^{2}}{(1+\beta_c)^{3}},
$$
$$
{\mathcal B} =  \frac {2\alpha^2_B I_n (g+\gamma M_c^2)^2}
{\gamma^2 x_c}+\frac {2\alpha^2_B I_n g (5x_c-3)(g+\gamma M_c^2)}
{\gamma^2 x_c(x_c-1)} 
$$
$$
-\frac{M_c^2(5x_c-3)}{x_c(\gamma-1)(x_c-1)}\frac{(\beta_c + \frac{3}{2\gamma})}{(1+\beta_c)} -\frac {8\alpha^2_B I_n g(g+\gamma M_c^2)}
{\gamma^2 (1+\beta_c)x_c}
$$
$$
 + \frac{2[3+\beta_c(\gamma+1)]M_c^2}{\gamma(\gamma-1)(1+\beta_c)^2 x_c} - \frac{M_c^2}{\gamma(\gamma-1)(1+\beta_c)(x_c-1)}
$$
$$
 - \frac{(4\zeta - 1)M_c^2}{2\gamma(\gamma-1)(1+\beta_c) x_c},
$$

$$
{\mathcal C} = -\frac {4\lambda_c \alpha_B I_n M_c (g+\gamma M_c^2)}{\gamma x_c^2},
$$

$$
{\mathcal D} = -\left[ \frac {\lambda_c^2}{x_c^3}-\frac {1}{2(x_c-1)^2}\right]
$$
$$
\times \left[\frac {[3+\beta_c(\gamma+1)]M_c^2}{(1+\beta_c)(\gamma-1)}-\frac{4\alpha^2_B g I_n (g+\gamma M_c^2)}{\gamma} \right].
$$
Here, the quantities with subscript `c' represent their values evaluated at the sonic point.

When a set of input parameters of the flow are provided, we can solve Eq. (15) to compute 
the sound speed at $x_c$. Subsequently, the radial velocity at $x_c$ can be 
obtained from Eq. (14). This allows us to investigate the properties of the sonic point 
using Eq. (13). At the sonic point, $d\upsilon/dx$ generally possesses two specific 
values: one corresponds to accretion flow and the other holds for wind solutions. If both 
the derivatives are real and of opposite sign, then the sonic point is called saddle type. Such 
a sonic point has a special significance in accretion flows as global transonic solutions only pass 
through it (Chakrabarti \& Das 2004). In the present paper, we focus to explore the 
dynamical structure and properties of global accretion flow, thus we leave the wind solutions aside 
for future study.

\section{Results and Discussions}

With the aim to obtain a global accretion solution, we solve equations corresponding to the gradients of 
$\upsilon$, $a$, $\lambda$ and $\beta$ simultaneously using known boundary values of angular momentum ($\lambda$), 
plasma $\beta$, accretion rate ($\dot{m}$) and $\alpha_B$ at a given radial distance 
($x$). Since black hole solutions must be transonic in nature, flow must pass through the 
sonic point. Hence, it is advantageous to supply the boundary values of the flow
at the sonic point. By this means, we integrate equations corresponding to the gradients of 
$\upsilon$, $a$, $\lambda$ and $\beta$ from the sonic point 
once inward up to just outside the black hole horizon and then outward up to a large distance 
(equivalently `outer edge of the disc'). These two parts can be joined to finally 
obtain a complete global transonic accretion solution. Depending on the input parameters, flow must 
possess at least one sonic point and presumably more (Chang \& Ostriker 1985; Lu {\em et al.} 1997; Das {\em et al.} 2001a). Sonic points which form close to the horizon, 
are called as inner sonic points ($x_{\rm in}$) and those that form far away from the horizon, are called
as outer sonic points ($x_{\rm out}$), respectively. In the sections that follow, we adopt 
$M_{\rm BH} = 10M_{\odot}$ as a fiducial value.

\subsection{Shock free global accretion solution}

\begin{figure}[!t]
\begin{center}
\includegraphics[width=1.0\columnwidth]{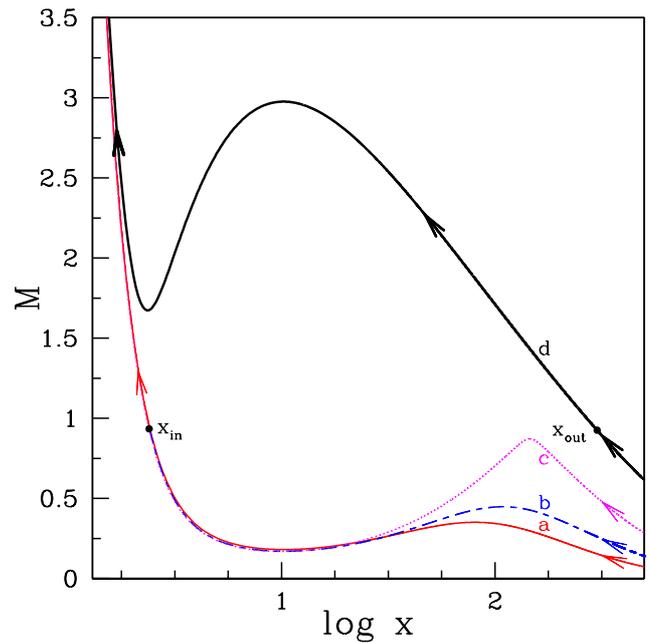}
\caption{Radial dependence of Mach number ($M=\upsilon/a$) of the accreting matter for
different values of angular momentum ($\lambda_{\rm edge}$) at the outer edge
$x_{\rm edge} = 500$ where $\beta_{\rm edge} = 3500$, $\alpha_B=0.023$ and 
$\dot{m} = 0.002$. Thin solid and short-long-dashed curves represent the results for 
$\lambda_{\rm edge}=5.6225$ and $3.6618$, respectively. For the same set of outer edge parameters, 
the minimum angular momentum that provides the accretion solution passing through
the inner sonic is identified as $\lambda^{\rm min}_{\rm edge}=2.6654$ (dotted curve).
When $\lambda_{\rm edge} < \lambda^{\rm min}_{\rm edge}$, accretion solutions pass through
the outer sonic point only (thick solid curve) where $\lambda_{\rm edge}=2.2183$.
In the figure, the locations of the inner sonic point ($x_{\rm in}$) and outer sonic
point ($x_{\rm out}$) are marked and arrows indicate the direction of the flow motion
towards the black hole. See text for details.}\label{figOne}
\end{center}
\end{figure}
In Fig. 1, we present variation of Mach number $(M = \upsilon/a)$ versus 
logarithmic radial distance ($x$). The solid curve, denoted by `a', depicts a 
global accretion solution passing through the inner sonic point $x_{\rm in} = 2.4242$ 
with angular momentum $\lambda_{\rm in} = 1.740$, $\beta_{\rm in} = 19.0$, 
$\alpha_B = 0.023$ and $\dot{m} = 0.002$, respectively, and connects the BH 
horizon with the outer edge of the disc $x_{\rm edge}$, where we note the values of the 
flow variables $\lambda_{\rm edge} = 5.6225$, $\beta_{\rm edge} = 3500$, 
$\upsilon_{\rm edge} = 0.00275$, $a_{\rm edge} = 0.03726$ at $x_{\rm edge} = 500$. 
In another way, we can obtain the same solution when the integration is carried out towards
the black hole starting from the outer edge of the disc ($x_{\rm edge}$) with
the noted boundary values. Hence, the above result necessarily describes the 
solution of an accretion flow that starts its journey from $x_{\rm edge} = 500$ and 
crosses the inner sonic point at $x_{\rm in} = 2.4242$ before crossing the 
event horizon. The arrow displays the direction of the flow. Now, we decrease 
$\lambda_{\rm edge} = 3.6618$ holding all the other values of the flow variables 
same at $x_{\rm edge} = 500$ and attain the global transonic solution by appropriately 
tuning the values of $\upsilon_{\rm edge} = 0.00467$ and $a_{\rm edge} = 0.03381$. The 
solution is denoted as `b'. Here, the values of $\upsilon_{\rm edge}$ and $a_{\rm edge}$ are 
further needed to start the integration as the sonic point is not identified a priori. 
Following this approach, we determine the minimum value of angular momentum at the 
outer edge $\lambda_{\rm edge}^{\rm min} = 2.6654$, below this value accretion solution 
fails to pass through the inner sonic point. Accretion solution which corresponds to the 
minimum $\lambda_{\rm edge}^{\rm min}$ is illustrated by the dotted curve and marked 
as `c'. The streamlines namely `a-c' depict solutions similar to the solution of 
advection dominated accretion flow around black holes (Narayan, Kato \& Honma 1997; 
Oda {\em et al.} 2007). However, yet another important class of solutions stay unexamined
which we deal with in this work. As $\lambda_{\rm edge}^{\rm min}$ is decreased further, such 
as $2.2183$, accretion solution changes its character and passes through the outer sonic point 
($x_{\rm out} = 300.89$) instead of inner sonic point ($x_{\rm in}$) with angular momentum 
$\lambda_{\rm out} = 2.0499$, $\beta_{\rm out} = 2283.02$ which is indicated by the thick 
solid line marked as `d'. In the context of magnetically supported accretion disc, there are only 
few studies of accretion solution passing through the outer sonic points (Sarkar \& Das 2015, 2016; 
Sarkar {\em {\em et al.}} 2018). Solutions specifically of this kind are appealing as they are likely to possess 
centrifugally supported shock waves. The existence of shock wave in an accretion flow has profound significance 
as it satisfactorily characterizes the spectral and temporal behaviour of numerous black hole sources 
(Molteni el al. 1999, Rao {\em et al.} 2000, Okuda {\em et al.} 2004, 2008, Das {\em et al.} 2014; Nandi {\em et al.} 2012; Aktar {\em et al.} 2015, 2017; Iyer, 
Nandi \& Mandal 2015, Sukov\'a \& Janiuk 2015, Sukov\'a {\em et al.} 2017). Thus, the present effort aims to explore the properties of magnetically supported 
accretion solutions that possess shock waves.

\subsection{Shock induced global accretion solution}

\begin{figure}[!t]
\begin{center}
\includegraphics[width=1.0\columnwidth]{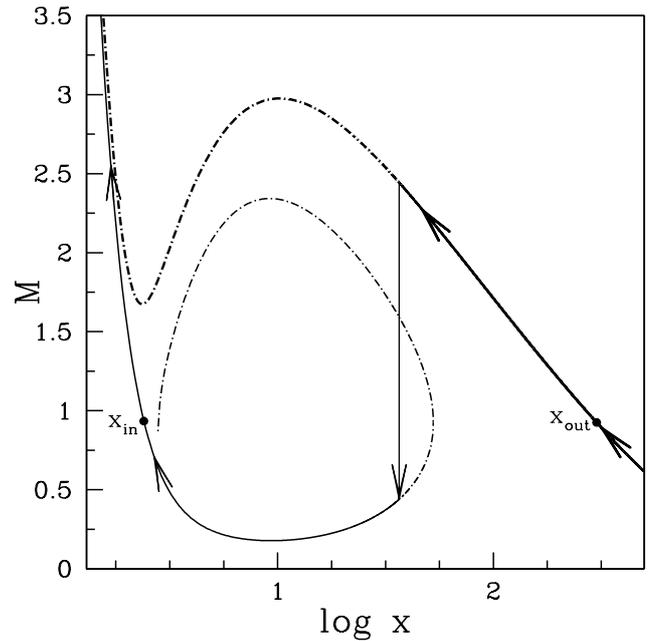}
\caption{A complete global accretion solution containing shock
($x_s = 36.60$) is depicted along with outer ($x_{\rm out}$) and inner ($x_{\rm in}$)
sonic points. Inflow parameters at the outer edge are same as case `d' of Fig. 1. See text
for details.}\label{figTwo}
\end{center}
\end{figure}
In Fig. 2, we present a composite global accretion solution containing shock wave where the flow 
transits the sonic location multiple times. Inflowing matter at the outer edge starts 
accreting towards the black hole subsonically with the boundary values same
as the case `d' of Fig. 1. The flow crosses the outer sonic point located at 
$x_{\rm out} = 300.89$ and becomes supersonic. When the rotating matter advances further, it experiences virtual 
barrier caused by the centrifugal repulsion and begins piling up there. The process persists until 
at some point, shock formation is triggered when the shock conditons are satisfied and this 
leads to a discontinuous transition of the flow variables. This is because when dynamically possible, shock 
is thermodynamically preferred accretion solution since the post-shock matter possesses high entropy content
(Becker \& Kazanas 2001). For a dissipative accretion flow, the expression of the entropy-accretion rate in 
the flow is obtained as (Chakrabarti 1996; Sarkar {\em {\em et al.}} 2018), 
$$
\dot{\cal{M}}(x) \propto \left(\frac{\beta}{1 + \beta}\right)^n a^{(2n+1)}\upsilon x^{3/2} (x - 1).
\eqno(16)
$$
In the absence of any dissipative processes like viscosity and/or radiative cooling, $\dot{\cal{M}}$ 
remains constant throughout the flow except at the shock. 

In an accretion flow, shock transition is governed by the conservation laws of mass, momentum, energy 
and magnetic field (Sarkar \& Das 2016, and reference therein). Across the shock
front, these laws are clearly given as the continuity of (a) mass flux (${\dot M}_{-}={\dot M_{+}}$) (b) the momentum flux 
($W_{-}+\Sigma_{-} \upsilon^2_{-}=W_{+}+\Sigma_{+} \upsilon^2_{+}$) (c) the energy flux 
(${\cal E_{-}}={\cal E_{+}}$) and (d) the magnetic flux ($\dot {\Phi}_{-}=\dot {\Phi}_{+}$) 
across the shock. Here, the quantities with subscripts `-' and `+' refer to values before and 
after the shock, respectively. While doing so, we inherently consider the shock to be thin and non-dissipative. 
The energy flux of the flow in the presence of a magnetic field is given by (Fukue 1990; Samadi {\em et al.} 2014),
$$
\mathcal{E} = \frac{\upsilon^2}{2} + \frac{a^2}{\gamma-1} + \frac{\lambda^2(x)}{2x^2} - \frac{1}{2(x-1)} + \frac{<B_\phi ^2>}{4\pi \rho}
\eqno(17) 
$$
In the immediate post-shock region, the flow becomes subsonic because of conversion of pre-shock 
kinetic energy in to the thermal energy. Hence, the post-shock matter becomes hot and dense. 
Subsonic post-shock matter continues its journey towards the BH because of gravitational attraction. It gains 
its radial velocity and consequently passes through the inner sonic point smoothly in order to satisfy the supersonic 
inner boundary condition before entering the event horizon. In Fig. 2, we present the variation of Mach number 
with the logarithmic radial distance. Thick curve refers to the accretion solution passing through the outer sonic 
point that can in principle enter in to the black hole directly. Interestingly, on the way towards the black hole, 
when the shock conditions are satisfied, flow makes discontinuous transition from the supersonic branch to the subsonic 
branch avoiding thick dot-dashed part of the solution. In the plot, the joining of the supersonic pre-shock flow with 
the subsonic post-shock flow is marked by the vertical arrow and the thin solid line denotes the inner part of the 
solution representing the post-shock flow. The overall direction of the flow motion during accretion 
towards black hole is indicated by the arrows.

\subsection{Shock dynamics and shock properties}

Here, we explore the dynamics of shock location as a function of the dissipation parameters 
($\beta_{\rm edge}$ and/or $\dot{m}$) for flows where initial parameters are held fixed. In Fig. 3, matter 
is injected subsonically at the outer edge of the disc, $x_{\rm edge} = 500$, with local specific energy, 
$\mathcal{E}_{\rm edge} = 9.4855 \times 10^{-4}$, local angular momentum, $\lambda_{\rm edge} = 2.2460$ 
and viscosity, $\alpha_B = 0.023$. When flow at $x_{\rm edge} = 500$ is injected with accretion rate 
$\dot{m} = 0.0001$ and plasma $\beta_{\rm edge} = 3500$, the flow suffers shock transition at $x_s = 246.62$. 
This result is represented in the figure by solid curve where vertical arrow represents the shock location. 
Next we increase the accretion rate to $\dot{m} = 0.004$, holding rest of flow parameters fixed at $x_{\rm edge}$ 
and notice that the shock front moves closer to the horizon at $x_s = 71.08$. This result is denoted using dotted 
curve where the dotted vertical arrow indicates the shock transition. Increase of $\dot{m}$ obviously augments the 
cooling rate of the flow. Since density and temperature of the flow undergo a catastrophic jump in the post-shock 
corona (PSC), effect of cooling in the PSC is intensified compared to the pre-shock flow. This reduces the post-shock 
thermal pressure and thus the shock moves closer to the black hole to maintain pressure balance across it. Additionally, 
we fix $\beta_{edge} = 200$ and $\dot{m} = 0.004$ maintaing the other flow parameters unchanged at $x_{\rm edge}$ and 
observe the shock to form at $x_s = 26.98$. We represent this result by dashed curve where dashed vertical line denotes 
the shock location as before. Decrease in $\beta_{edge}$ indicates the increase of magnetic fields in the 
accretion flow that augments the Maxwell stress, thus enhancing the angular momentum transport from the inner to the 
outer region in the disc. Thus the centrifugal repulsion at the PSC weakens. Furthermore, decrease of $\beta_{\rm edge}$ 
ultimately enhances the synchrotron cooling efficiency as well. Consequently, the collective effects of both the physical 
processes drive the shock front even further towards the horizon. This highlights the important role played by 
$\beta_{edge}$ in deciding the dynamics of shock location, apart from the role played by $\dot{m}$.

\begin{figure}[!t]
\begin{center}
\includegraphics[width=1.0\columnwidth]{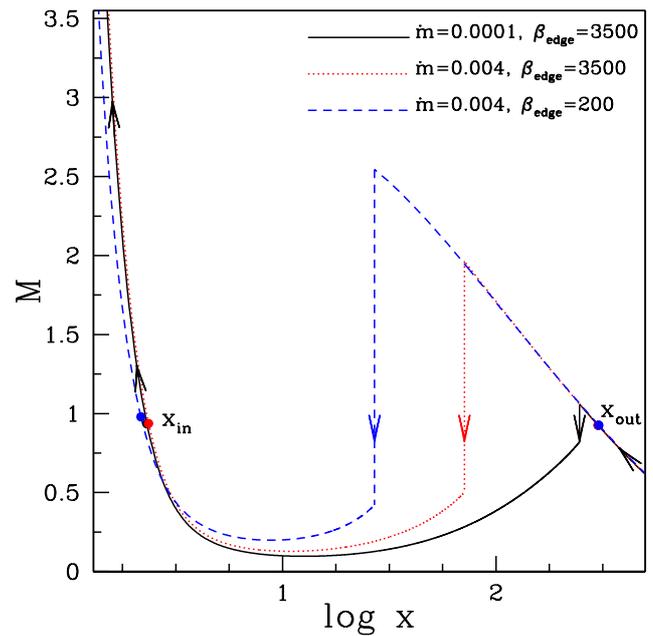}
\caption{Variation of Mach number as function of logarithmic radial distance. Flows are injected from the outer edge
$x_{\rm edge} = 500$ with energy $\cal{E}_{\rm edge}$ = $9.4855 \times 10^{-4}$, angular momentum $\lambda_{\rm edge} = 2.2460$ and viscosity $\alpha_B = 0.023$. Solid, dotted
and dashed curves depict the results obtained for ($\dot{m}, \beta_{\rm edge})$ = $(10^{-4}, 3500), (0.004, 3500)$ and $(0.004, 200)$, respectively. Vertical arrows indicate the corresponding shock transitions positioned at $x_s = 246.62$ (solid), $x_s$ = 71.08 (dotted) and $x_s$ = 26.98 (dashed). See text for details.
}\label{figThree}
\end{center}
\end{figure}
\begin{figure}[!t]
\begin{center}
\includegraphics[width=1.0\columnwidth]{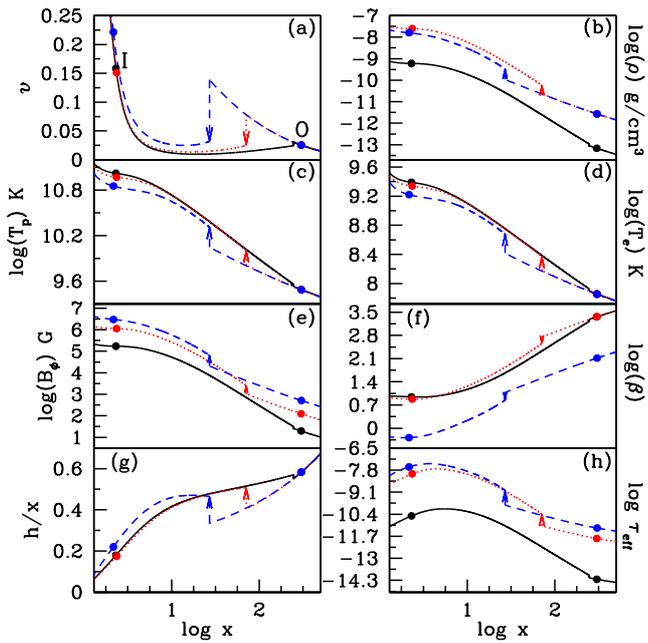}
\caption{Variation of (a) radial velocity, (b) density in $g/cm^3$ ,
(c) proton temperature, (d) electron temperature, (e) strength of magnetic field, 
(f) ratio of gas pressure to magnetic pressure, (g) disc scale height ($h/x$) 
and (h) effective optical depth as function of logarithmic radial coordinate. Results 
plotted with solid, dotted and dashed curves correspond to the accretion solution
depicted in Fig. 3. Filled circles represent the sonic points where
the closer one is the inner sonic point and the furthest one is the
outer sonic point. Vertical arrows indicate the shock position. See
text for details
}\label{figFour}
\end{center}
\end{figure}
In Fig. 4, we demonstrate the vertically averaged accretion disc structure corresponding to the 
solutions considered in Fig. 3. In every panel, we depict the variation of a flow variable with 
logarithmic radial distance, where vertical arrows point out the shock transition. In Fig. 4(a), we plot the radial velocity 
profile ($\upsilon$) of the accretion flow. As anticipated, $\upsilon$ increases with the decrease of 
radial coordinate before it experiences a shock transition. After the shock transition $\upsilon$ drops to 
a subsonic value and again increases steadily in the PSC. Ultimately, flow enters the event horizon with 
velocity of the order of speed of light after crossing the inner sonic point. Here, solid, dotted and dashed 
curves exhibit results for ($\dot{m}$, $\beta_{\rm edge}$) = ($10^{-4}$, $3500$), ($0.004$, $3500$) and 
($0.004$, $200$), respectively. We demonstrate the density profile of the flow in Fig. 4(b), 
where, we find an increase of density directly after the shock transition in each case. This occurs due 
to the decline of radial velocity at PSC, eventually preserving the conservation of mass flux across the shock front. 
The general trend of density profile corresponding to $\dot{m} = 0.004$ is higher compared to the case of $\dot{m} = 10^{-4}$ 
just because the large $\dot{m}$ signifies higher mass inflow at the outer edge. In Fig. 4(c), the proton temperature 
profile ($T_p$) is shown. During the shock transition, supersonic pre-shock flow is turned in to subsonic flow, and thereby, 
greater part of the kinetic energy of the infalling matter is converted to the thermal energy at PSC. This ultimately causes heating 
of the PSC as revealed by the rise of post-shock temperature profile. Additionally, when $\dot{m}$ rises, radiative cooling is 
more effective at PSC that explains the reduction of $T_p$ as clearly visible in the vicinity of the event horizon. Further, 
we notice that the decrease of $\beta_{\rm edge}$ necessarily displays the decrease of temperature profile of the flow.
Fig. 4(d) shows the variation of electron temperature ($T_e$) with logarithmic radial distance. Since we have assumed the electron 
temperature to be estimated using proton temperature as $T_e\sim0.023T_p$, the radial variation of $T_e$ follows the same trend as $T_p$. 
Importantly, since the radiative cooling is mostly effective at the inner part of the disc (due to high density, temperature and 
magnetic field), the synchrotron cooling is seen to be significantly effective in this region leading to decrease in $T_e$ 
profile. In Fig. 4(e), we show the variation of toroidal magnetic field ($B_\phi$) with logarithmic radial distance. 
$B_\phi$ increases radially as the flow approaches the BH because of increase of magnetic flux as matter approaches 
the event horizon. Compression of flow in the post-shock region causes the magnetic field strength to increase after 
the shock because of the conservation of magnetic flux ($\dot{\Phi}$) across the shock. Increase of accretion rate in 
the flow increases $B_\phi$ (through gas pressure) which in turn enhances angular momentum transport and 
pushes the shock front towards the BH. We present the radial variation 
of plasma $\beta$ profile in Fig. 4(f). We observe that $\beta$ decreases as the flow approaches to the horizon. 
Furthermore, $\beta$ drops sharply across the shock that renders the PSC into magnetically dominated. The radial 
variation of the vertical scale-height ($h/x$) is presented in Fig. 4(g). The validity of the thin disc approximation 
is found to hold all throughout from the outer edge to the horizon, in spite of the presence of shock transition. Lastly, 
in Fig. 4(h), we plot the variation of effective vertical optical depth, $\tau_{\rm eff} = \sqrt{\tau_{\rm es}\tau_{\rm syn}}$ 
(Rajesh \& Mukhopadhyay 2010) where, $\tau_{\rm es}$ represents the scattering optical depth given by $\tau_{\rm es} = \kappa_{\rm es} \rho h$ 
and the electron scattering opacity, $\kappa_{\rm es}$, is taken to be $\kappa_{\rm es} = 0.38~{\rm cm}^2 {\rm g}^{-1}$.
Here, $\tau_{\rm syn}$ denotes the absorption effect arising due to thermal processes and is given by 
$\tau_{\rm syn} =\left( h q_{\rm syn}/4 \sigma T_{e}^4\right)\left(2GM_{\rm BH}/c^2\right)$ (Rajesh \& Mukhopadhyay 2010)
where, $q_{\rm syn}$ is the synchrotron emissivity (Shapiro \& Teukolsky 1983) and $\sigma$ is the Stefan-Boltzmann 
constant. We observe that the optical depth of PSC is consistently greater than the pre-shock region because the 
density of post-shock region is higher (see, Fig. 4b). Moreover, the global variation of the optical depth for 
augmented accretion rate stays higher. In addition, despite of a steep density profile, PSC flow is found 
to remain optically thin ($\tau < 1$). This instinctively leads to the conclusion that there is a significant 
possibility of escaping hard radiations from the PSC.

Fig. 5, demonstrates the different shock properties as a function of accretion rate ($\dot{m}$) for 
flows injected from a fixed outer edge as $x_{\rm edge} = 500$ with $\beta_{\rm edge}$ = 3500, 
energy $\cal{E}_{\rm edge}$ = $9.4855 \times 10^{-4}$ and viscosity $\alpha_B$ = 0.023. In 
the upper panel (Fig. 5a), the variation of shock position is shown for three different values 
of angular momentum ($\lambda_{\rm edge}$) at $x_{\rm edge}$. The solid, dotted and dashed curves 
correspond to flows injected with angular momentum $\lambda_{\rm edge}$ = $2.2120$, $2.2290$ 
and $2.2460$ respectively. Evidently, we find that a wide range of $\dot{m}$ supports shock induced 
global accretion solutions. For a given $\lambda_{\rm edge}$, the location of shock advances towards 
the horizon with the increment of accretion rate ($\dot{m}$). With the rise of accretion rate the 
efficiency of radiative cooling is enhanced and the flow loses energy during accretion. Loss of energy causes 
drop in post-shock thermal pressure thus pushing the shock front closer to the horizon in order 
to maintain pressure balance across the shock. When accretion rate exceeds its critical value 
($\dot{m}^{\rm cri}$), standing shock no longer form as the shock conditions are not fulfilled. 
This clearly suggests that the likelihood of shock formation reduces with the increase 
of $\dot{m}$. It should be noted that ($\dot{m}^{\rm cri}$) does not have a global value, rather it 
primarily depends on the accretion flow parameters. Interestingly enough, when $\dot{m} > (\dot{m}^{\rm cri})$, 
flow can have oscillatory shocks, however, the investigation of non-steady shock properties are out of 
scope of the present work. Such type of oscillation of the post-shock region has been shown by Molteni {\em et al.} (1996). 
In addition, for a given $\dot{m}$, shock front recedes from the horizon 
with the increase of $\lambda_{\rm edge}$. This is because the centrifugal barrier strengthens with the 
increase of $\lambda_{\rm edge}$. From this it is clear that shocks in accretion flow around black 
holes are centrifugally driven. It is evident from Fig. 5(a) that when $\lambda_{\rm edge}$ is decreased 
the range of $\dot{m}$ over which shock exists, decreases. When $\lambda_{\rm edge}$ will be decreased 
below the value $2.2120$, shocks will exist for still lower range of $\dot{m}$ and will finally cease to 
exist when $\lambda_{\rm edge}$ will be decreased beyond a critical limit because standing shock will fail 
to form. It is previously noted that the shock 
\begin{figure}[!t]
\begin{center}
\includegraphics[width=1.0\columnwidth]{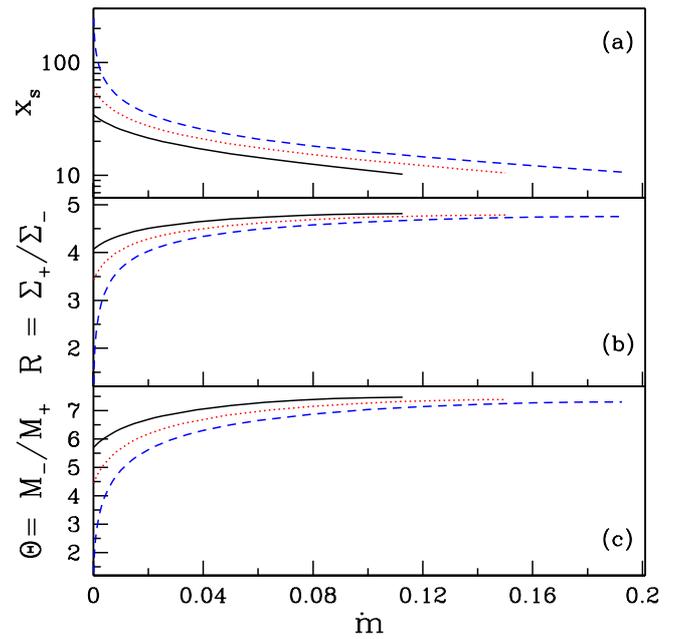}
\caption{Variation of (a) shock location $x_s$, (b) compression
ratio $R$, and (c) shock strength $\Theta$ as function of $\dot{m}$ for flows
injected from $x_{\rm edge}$ = 500 with $\alpha_B$ = 0.023, $\beta_{\rm edge} = 3500$ and
$\cal{E}_{\rm edge}$ = $9.4855 \times 10^{-4}$. Solid, dotted and dashed curves represent
the results corresponding to $\lambda_{\rm edge}$ = $2.2120$, $2.2290$ and $2.2460$, 
respectively. See text for details.
}\label{figFive}
\end{center}
\end{figure}
compression causes the density and temperature of the PSC 
to increase. Furthermore, the spectral properties of an accretion disc are directly determined by the 
density and temperature distribution of the flow. Hence, it is imperative to estimate the amount of density 
and temperature boost across the shock transition. For this, we first evaluate the compression 
ratio which is defined as the ratio of the vertically averaged post-shock density to the pre-shock 
density ($R = \Sigma_+ / \Sigma_-$) and plot it as function of $\dot{m}$ in Fig. 5b. We retain the flow 
parameters chosen in Fig. 5a. For a given $\lambda_{\rm edge}$, $R$ is seen to grow monotonicaly with the increase 
of $\dot{m}$. That is so because the shock is pushed closer to the horizon with the increase of $\dot{m}$ 
boosting the density compression and resulting in increase of compression ratio. Oppositely, for a given $\dot{m}$, 
when $\lambda_{\rm edge}$ is increased, the centrifugal barrier strengthens and shock recedes away form the horizon 
leading to decrease in the post-shock compression. When $\dot{m} > (\dot{m}^{\rm cri})$, since shock 
cease to exist, we observe a cut-off in the compression ratio in all the cases. Next, we compute the shock 
strength ($\Theta$) defined as the ratio of pre-shock Mach number ($M_{-}$) to the post-shock Mach number 
($M_{+}$) which basically measures the temperature jump across the shock. In Fig. 5(c), we plot the variation 
of $\Theta$ as a function of accretion rate $\dot{m}$ for the same injection parameters as in Fig. 5(a). We 
observe that the response of $\Theta$ on the increase of $\dot{m}$ is identical to $R$ as represented in 
Fig. 5(b).

\section{Conclusions}
In this work, we demonstrate that the synchrotron cooling plays an important role 
in a magnetized accretion disc around black holes. Efforts are still ongoing to study the origin of viscosity 
and the exact mode of angular momentum transport in accretion discs by several authors (Menou 2000, Balbus 2003, Becker \& Subramanian 2005, Kaufmann {\em et al.} 2007, King {\em et al.} 2007, Guan \& Gammie 2009, Hopkins \& Quataert 2011, Kotko \& Lasota 2012). 
Hence, we rely on the numerical simulation results of Machida {\em et al.} (2006). 
Following Machida {\em et al.} (2006), we assume the x$\phi$ -component of the Maxwell stress to be
proportional to the total pressure. We find that such a flow is transonic and 
passes through the sonic point multiple times giving rise
to the phenomenon of shocks in the accretion flow. Global accretion
solutions of magnetically supported accretion discs around stationary black
 holes have been carried out by Oda {\em et al.} (2007, 2012).
These works show the accretion solutions passing through a
single sonic point only and ignore the possibility of multiple sonic points
in the flow. Thus the solutions presented by Oda {\em et al.} (2007, 2012)
are a subset of the generalized accretion solutions studied in this
work. In Sarkar \& Das (2016), the magnetic field strength was assumed to 
be moderate throughout the flow. However the present
work has no such restrictions and shock solutions have been shown
to be present in the disc when the inner region is magnetically dominated ($\beta < 1$) as well (Fig. 4f). 
We make a detailed study of the effects of the dissipation variables, namely accretion 
rate ($\dot{m}$) and magnetic parameter $\beta$ on the global accretion solutions. We find that 
increase in radiative loss due to increase of cooling and strength 
of magnetic field removes a significant thermal pressure from the inflow and
as the dissipation increases, the shock location shifts towards the black hole to maintain 
the pressure balance across it. When the dissipation is reached its critical value, standing 
shocks can no longer form. Such a study is highly important since the dissipation parameters 
are likely to influence the spectral and timing properties of the
radiation emitted from the disc. We have studied the effect of cooling on the shock dynamics. 
For a flow strating from a large distance with a given set of outer edge conditions, there exists 
a cut-off in accretion rate, beyond which, steady shock solutions cannot exist, as in Fig. 5. 
Beyond the critical limit, unsteady, oscillating shocks are possible (Das {\em et al.} 2014).
We ignore the formation of such oscillating shocks since their study is out of the scope of 
the present work.

Finally, we point out the limitations of the present work. In
this paper, the adiabatic index has been considered to be constant
throughout the flow instead of calculating it self-consistently. Also,
the black hole in this paper has been considered to be non-rotating.
Apart from plasma $\beta$, the spin of the black hole is also expected to
affect the dynamics of shocks. We would like to consider the above
issues in the future works.








\begin{thebibliography}{99} 
\bibitem{latexcompanion} 
Aktar, R., Das, S., Nandi, A. 2015, {\em MNRAS}, {\bf 453}, 3414.
\bibitem{latexcompanion} 
Aktar, R., Das, S., Nandi, A., Sreehari, H. 2017, {\em MNRAS}, {\bf 471}, 4806.
\bibitem{latexcompanion} 
Balbus, S., Hawley, J. F. 1991, {\em ApJ}, {\bf 376}, 214.
\bibitem{latexcompanion}
Balbus, S. 2003, {\em ARA}\&{\em A}, {\bf 41}, 555.
\bibitem{latexcompanion} 
Becker, P. A., Kazanas, D. 2001, {\em ApJ}, {\bf 546}, 429.
\bibitem{latexcompanion}
Becker, P. A., Subramanian, P. 2005, {\em ApJ}, {\bf 622}, 520.
\bibitem{latexcompanion}
Beckwith, K., Hawley, J. F., Krolik, J. H. 2008, {\em ApJ}, {\bf 678}, 1180.
\bibitem{latexcompanion}
Blandford, R. D., Znajek, R. L. 1977, {\em MNRAS}, {\bf 179}, 433.
\bibitem{latexcompanion} 
Chakrabarti, S. K. 1989, {\em ApJ}, {\bf 347}, 365.
\bibitem{latexcompanion} 
Chakrabarti, S. K. 1990, {\em MNRAS}, {\bf 243}, 610.
\bibitem{latexcompanion} 
Chakrabarti, S. K. 1996, {\em ApJ}, {\bf 464}, 664.
\bibitem{latexcompanion}
Chakrabarti, S. K. 1999, {\em A}\&{\em A}, {\bf 351}, 185
\bibitem{latexcompanion} 
Chakrabarti, S. K., Das, S. 2004, {\em MNRAS}, {\bf 349}, 649.
\bibitem{latexcompanion}
Chang, K. M., Ostriker, J. P. 1985, {\em ApJ}, {\bf 288}, 428.
\bibitem{latexcompanion} 
Chattopadhyay, I., Chakrabarti, S. K. 2002, {\em MNRAS}, {\bf 333}, 454.
\bibitem{latexcompanion} 
Chattopadhyay, I., Chakrabarti, S. K. 2011, {\em IJMPD}, {\bf 20}, 1597.
\bibitem{latexcompanion} 
Das, S., Chattopadhyay, I., Chakrabarti, S. K. 2001a, {\em ApJ}, {\bf 557}, 983.
\bibitem{latexcompanion} 
Das, S., Chattopadhyay, I., Nandi, A., Chakrabarti, S. K. 2001b, {\em A}\&{\em A}, {\bf 379}, 683.
\bibitem{latexcompanion} 
Das, S. 2007, {\em MNRAS}, {\bf 376}, 1659.
\bibitem{latexcompanion} 
Das, S., Chattopadhyay, I., Nandi, A., Molteni, D. 2014, {\em MNRAS}, {\bf 442}, 251.
\bibitem{latexcompanion}
De Villiers, J.-P., Hawley, J. F., Krolik, J. H., Hirose, S. 2005, {\em ApJ}, {\bf 620}, 878.
\bibitem{latexcompanion} 
Fukue, J. 1987, {\em PASJ}, {\bf 39}, 309.
\bibitem{latexcompanion} 
Fukue, J. 1990, {\em PASJ}, {\bf 42}, 793.
\bibitem{latexcompanion}
Fukumura K., Tsuruta S. 2004, {\em ApJ}, {\bf 611}, 964.
\bibitem{latexcompanion}
Gierli\'nski, M., Newton, J. 2006, {\em MNRAS}, {\bf 370}, 837.
\bibitem{latexcompanion}
Guan, X., Gammie, C. F. 2009, {\em ApJ}, {\bf 697}, 1901.
\bibitem{latexcompanion} 
Hawley, J. F. 2000, {\em ApJ}, {\bf 528}, 462.
\bibitem{latexcompanion}
Hawley, J. F. 2001, {\em ApJ}, {\bf 554}, 534.
\bibitem{latexcompanion} 
Hawley, J. F., Krolik, J. H. 2001, {\em ApJ}, {\bf 548}, 348.
\bibitem{latexcompanion}
Hawley, J. F., Krolik, J. H. 2002, {\em ApJ}, {\bf 566}, 164.
\bibitem{latexcompanion}
Hirose, S., Krolik, J. H., De Villiers, J. P., Hawley, J. F. 2004, {\em ApJ}, {\bf 606}, 1083.
\bibitem{latexcompanion} 
Hirose, S., Krolik, J. H., Stone, J. M. 2006, {\em ApJ}, {\bf 640}, 901.
\bibitem{latexcompanion}
Hopkins P. F., Quataert E. 2011, {\em MNRAS}, {\bf 415}, 1027.
\bibitem{latexcompanion} 
Ichimaru, S. 1977, {\em ApJ}, {\bf 214}, 840.
\bibitem{latexcompanion}
Igumenshchev, I. V., Narayan, R., \& Abramowicz, M. A. 2003, {\em ApJ}, {\bf 592}, 1042.
\bibitem{latexcompanion} 
Iyer, N., Nandi, A., Mandal, S. 2015, {\em ApJ}, {\bf 807}, 108.
\bibitem{latexcompanion} 
Johansen, A., Levin, Y. 2008, {\em A}\&{\em A}, {\bf 490}, 501.
\bibitem{latexcompanion} 
McKinney, J. C., Gammie, C. F. 2004, {\em ApJ}, {\bf 611}, 977.
\bibitem{latexcompanion} 
Kato, Y., Mineshige, S., Shibata, K. 2004, {\em ApJ}, {\bf 605}, 307.
\bibitem{latexcompanion} 
Kaufmann T., Mayer L., Wadsley J., Stadel J. and Moore B. 2007, {\em MNRAS}, {\bf 375}, 53.
\bibitem{latexcompanion}
King A. R., Pringle J. E. \& Livio M. 2007, {\em MNRAS}, {\bf 376}, 1740.
\bibitem{latexcompanion}
Koide S., Shibata K., Kudoh T., Meier D. L. 2002, {\em Science}, {\bf 295}, 1688.
\bibitem{latexcompanion}
Kotko, I., Lasota, J.-P. 2012, {\em A}\&{\em A}, {\bf 545}, A115.
\bibitem{latexcompanion} 
Krolik, J. H., Hirose, S., Blaes, O. 2007, {\em ApJ}, {\bf 664}, 1045.
\bibitem{latexcompanion} 
Landau L. D., Lifshitz E. D. 1959, {\em Fluid Mechanics} (New York, Pergamon).
\bibitem{latexcompanion} 
Lebedev, S. V., Ciardi, A., Ampleford, D. J., {\em et al.} 2005, {\em MNRAS}, {\bf 361}, 97.
\bibitem{latexcompanion}
Lu, J. F., Yu, K. N., Yuan, F., Young, E. C. M. 1997, {\em A}\&{\em A}, {\bf 321}, 665.
\bibitem{latexcompanion}
Lu, J. F., Yuan, F. 1998, {\em MNRAS}, {\bf 295}, 66.
\bibitem{latexcompanion} 
Machida, M., Matsumoto, R. 2003, {\em ApJ}, {\bf 585}, 429.
\bibitem{latexcompanion} 
Machida, M., Nakamura, K. E., Matsumoto, R. 2006, {\em PASJ}, {\bf 58}, 193.
\bibitem{latexcompanion} 
Mandal, S., Chakrabarti, S. K. 2005, {\em Ap}\&{\em SS}, {\bf 297}, 269.
\bibitem{latexcompanion} 
Matsumoto, R., Kato, S., Fukue, J., Okazaki, A. T. 1984, {\em PASJ}, {\bf 36}, 71.
\bibitem{latexcompanion} 
McKinney, J., Blandford, R. 2009, {\em MNRAS}, {\bf 394}, L126.
\bibitem{latexcompanion}
Menou, K. 2000, {\em Sci}, {\bf 288}, 2022.
\bibitem{latexcompanion}
Molteni, D., Sponholtz, H., Chakrabarti, S. K. 1996, {\em ApJ}, {\bf 457}, 805.
\bibitem{latexcompanion}
Molteni D., Toth G., Kuznetsov O. A. 1999, {\em ApJ}, {\bf 516}, 411.
\bibitem{latexcompanion} 
Narayan, R., Yi, I. 1995, {\em ApJ}, {\bf 452}, 710.
\bibitem{latexcompanion} 
Narayan, R., Kato, S., Honma, F. 1997, {\em ApJ}, {\bf 476}, 49.
\bibitem{latexcompanion} 
Nandi, A., Debnath, D., Mandal, S., Chakrabarti, S. K. 2012, {\em A}\&{\em A}, {\bf 542}, 56.
\bibitem{latexcompanion} 
Oda, H., Machida, M., Nakamura, K. E., Matsumoto, R. 2007, {\em PASJ}, {\bf 59}, 457.
\bibitem{latexcompanion} 
Oda, H., Machida, M., Nakamura, K. E., Matsumoto, R. 2010, {\em ApJ}, {\bf 712}, 639.
\bibitem{latexcompanion} 
Oda, H., Machida, M., Nakamura, K. E., Matsumoto, R., Narayan, R. 2012, {\em PASJ}, {\bf 64}, 15.
\bibitem{latexcompanion} 
Okuda T., Teresi V., Toscano E., Molteni D. 2004, {\em PASJ}, {\bf 56}, 547.
\bibitem{latexcompanion} 	
Okuda, T., Teresi, V., Molteni, D. 2008, AIP Conference Proceedings, vol. 968, p. 417.
\bibitem{latexcompanion} 
Paczy\'nski, B., Wiita, P. J. 1980, {\em A}\&{\em A}, {\bf 88}, 23.
\bibitem{latexcompanion}
Papaloizou, J. C. B., Terquem, C. 1997, {\em MNRAS}, {\bf 287}, 771.
\bibitem{latexcompanion} 
Parker, E. N. 1966, {\em ApJ}, {\bf 145}, 811.
\bibitem{latexcompanion}
Pudritz, R. E., Norman, C. A. 1986, {\em Can. J. Phys.}, {\bf 64}, 501.
\bibitem{latexcompanion}
Rajesh, S. R., Mukhopadhyay B. 2010, {\em MNRAS}, {\bf 402}, 961.
\bibitem{latexcompanion}
Rao A. R., Yadav J. S., Paul B. 2000, {\em ApJ}, {\bf 544}, 443.
\bibitem{latexcompanion} 
Samadi, M., Abbassi, S., Khajavi, M. 2014, {\em MNRAS}, {\bf 437}, 3124.
\bibitem{latexcompanion} 
Sarkar, B., Das, S. 2015, {\em ASInC}, {\bf 12}, 91.
\bibitem{latexcompanion} 
Sarkar, B., Das, S. 2016, {\em MNRAS}, {\bf 461}, 190.
\bibitem{latexcompanion} 
Sarkar, B., Das, S., Mandal, S. 2018, {\em MNRAS}, {\bf 473}, 2415.
\bibitem{latexcompanion} 
Shakura, N. I., Sunyaev, R. A. 1973, {\em A}\&{\em A}, {\bf 24}, 337.
\bibitem{latexcompanion} 
Shapiro, S. L., Teukolsky, S. A. 1983, {\em Black Holes, White Dwarfs and Neutron Stars: The Physics of Compact Objects} (New York, Wiley).
\bibitem{latexcompanion} 
Shibata, K., Tajima, T., \& Matsumoto, R. 1990, {\em ApJ}, {\bf 350}, 295.
\bibitem{latexcompanion} 
Singh, C. B., Chakrabarti, S. K. 2011, {\em MNRAS}, {\bf 410}, 2414.
\bibitem{latexcompanion} 
Spruit, H. C., \& Uzdensky, D. A. 2005, {\em ApJ}, {\bf 629}, 960.
\bibitem{latexcompanion} 
Sukov\'a P., Janiuk A. 2015, {\em MNRAS}, {\bf 447}, 1565.
\bibitem{latexcompanion} 
Sukov\'a P., Charzy\'nski, S., Janiuk A. 2017, {\em MNRAS}, {\bf 472}, 4327.
\bibitem{latexcompanion} 
Ustyugova, G. V., Koldoba, A. V., Romanova, M. M., Chechetkin, V. M., \& Lovelace, R. V. E. 1999, {\em ApJ}, {\bf 516}, 221.
\bibitem{latexcompanion} 
Yuan, F. 2001, {\em MNRAS}, {\bf 324}, 119.

\end{thebibliography}
\end{document}